\documentclass[preprint,showpacs,preprintnumbers,amsmath,amssymb,floatfix,endfloats*]{revtex4}

\usepackage{graphicx}
\usepackage{dcolumn}
\usepackage{bm}
\usepackage{amsfonts}

\DeclareMathAlphabet{\mathsfsl}{OT1}{cmr}{bx}{it}
\begin{document}
\title{Molecular diffusion and slip boundary conditions at smooth surfaces with periodic and random nanoscale textures}
\author{Nikolai V. Priezjev}
\affiliation{Department of Mechanical Engineering, Michigan State
University, East Lansing, Michigan 48824}
\date{\today}
%
\begin{abstract}

The influence of periodic and random surface textures on the flow
structure and effective slip length in Newtonian fluids is
investigated by molecular dynamics (MD) simulations.  We consider a
situation where the typical pattern size is smaller than the channel
height and the local boundary conditions at wetting and nonwetting
regions are characterized by finite slip lengths.  In case of
anisotropic patterns, transverse flow profiles are reported for
flows over alternating stripes of different wettability when the
shear flow direction is misaligned with respect to the stripe
orientation. The angular dependence of the effective slip length
obtained from MD simulations is in good agreement with hydrodynamic
predictions provided that the stripe width is larger than several
molecular diameters.  We found that the longitudinal component of
the slip velocity along the shear flow direction is proportional to
the interfacial diffusion coefficient of fluid monomers in that
direction at equilibrium. In case of random textures, the effective
slip length and the diffusion coefficient of fluid monomers in the
first layer near the heterogeneous surface depend sensitively on the
total area of wetting regions.

\end{abstract}



\maketitle

\section{Introduction}

Modeling fluid flows over chemically or topographically patterned
substrates is important for micro and nanofluidic applications
involving mixing~\cite{Stroock02Chem} and separation
processes~\cite{Squires05}.  As the surface to volume ratio
increases, the role of hydrodynamic boundary conditions in
determining fluid velocity profiles becomes dominant.  It is well
recognized now that the classical no-slip boundary condition can, in
principle, be violated and the velocity profiles can be
significantly affected by the interfacial slip~\cite{CharlaixRev10}.
The degree of slip is usually quantified in terms of the Navier slip
length, which is defined as a distance between the real interface
and imaginary plane where the extrapolated tangential velocity
component vanishes. The magnitude of the slip length for smooth
nonwetting surfaces is typically on the order of tens of
nanometers~\cite{Churaev84,Charlaix05,Vinograd06,Yoda10,AttardLang11};
however, in special cases of nanoengineered superhydrophobic
surfaces slip lengths in the micron range were
reported~\cite{Rothstein04,Forest06,Vorobieff06,Kim09}. Although the
Navier-Stokes equation with slip boundary conditions is often used
to model small scale flows, the limits of validity of the continuum
description of complex flows at nanometer scales remain not fully
understood~\cite{CharlaixRev10}.

In recent years, a number of molecular dynamics (MD) studies have
examined factors that determine the magnitude of the slip length at
interfaces between crystalline surfaces and monatomic
liquids~\cite{Fischer89,KB89,Thompson90,Nature97,Barrat99,Barrat99fd,Travis00,Quirke01,Attard04,Priezjev07,PriezjevJCP,Tzeng07,Li09,Asproulis10,Freund11}.
One of the most important conclusions is that the degree of slip
strongly correlates with the in-plane structure in the first fluid
layer induced by the periodic surface potential~\cite{Thompson90}.
An estimate of the low-shear-rate limit of the slip length can be
obtained via the Green-Kubo relation between the friction
coefficient at the interface and the time integral of the
autocorrelation function of the lateral force that acts on the
adjacent fluid from the solid wall~\cite{Barrat99fd}. In general,
when the surface energy is weak, the slip length is constant only at
relatively low shear rates and it increases nonlinearly at high
shear rates~\cite{Nature97}. It was recently demonstrated that the
linear regime of slip holds when the slip velocity of the first
fluid layer is smaller than the diffusion rate of fluid monomers
over the distance between the nearest minima of the periodic surface
potential at equilibrium~\cite{Priezjev10}. It was also found that
at sufficiently high shear rates, the slip length becomes
anisotropic for dense walls with weak surface energy; and, in
particular, the slip length increases when the flow is oriented
along the crystallographic axis of the wall
lattice~\cite{Priezjev10}.

Several studies considered the flow of a Newtonian fluid over
surfaces patterned with stripes of different wettability using both
molecular dynamics and continuum
simulations~\cite{Bocquet04,Priezjev05,Sheng05,Hendy05}. In the
presence of flow, a heterogeneous surface with mixed boundary
conditions induces spatial variations in the velocity profiles. The
flow profiles averaged on length scales larger than the typical
pattern size can be used to define the \textit{effective slip
length}, which describes the flow away from the surface.   The
comparative analysis between MD and continuum simulations
demonstrated that there is an excellent agreement between the
velocity profiles and effective slip lengths for stripe widths
larger than approximately 30 molecular diameters for flow
configurations either parallel or perpendicular to the stripe
orientation~\cite{Priezjev05}. Later studies have shown that similar
conclusions hold for slip flows of
Newtonian~\cite{Priezjev06,Niavarani10} and
polymeric~\cite{Niavarani08} fluids over periodically corrugated
surfaces. In a more general situation, when the mean flow direction
is not aligned with the symmetry axis of surface patterns, it is
expected that the slip velocity will acquire a non-zero transverse
component. In MD simulations, the transverse velocity profiles were
reported in force-driven flows over flat surfaces with asymmetric
distribution of wetting regions~\cite{Hendy05} and in spiral flows
inside a cylindrical channel~\cite{Korea11}.   One of the goals of
the present study is to perform a detailed comparative analysis of
the effective slip length and velocity fields in flows around
anisotropic textured surfaces using MD simulations and recent
analytical results~\cite{Wang03,Teo09,Vinograd10}.

The Navier slip boundary condition for flows over arbitrary
patterned surfaces can be formulated in tensor form, i.e., the
apparent slip velocity vector is equal to the product of normal
traction and an interfacial mobility
tensor~\cite{Vinograd08,Kamrin10,Vinograd11}. It was suggested that
at the microscopic level, the mobility tensor is related to the
interfacial diffusivity per unit area~\cite{Vinograd08}.  Recently,
it was also proven that for steady noninertial flows over surfaces
perturbed by arbitrary periodic height and local slip fluctuations
the mobility tensor is always symmetric~\cite{Stone11}.  Using the
theory of transport in heterogeneous media, rigorous bounds on the
effective slip length were obtained for arbitrary two-component
texture with given area fractions and local slip
lengths~\cite{Vinograd09,Vinograd10R}. In particular, it was shown
that parallel (perpendicular) stripe orientation with respect to the
mean flow direction results in maximum (minimum) slip flow in a thin
channel~\cite{Vinograd09}. In case when the mean flow is not
parallel or perpendicular to stripes, the angular dependence of the
effective slip length was derived
analytically~\cite{Wang03,Vinograd08} and later verified by lattice
Boltzmann simulations~\cite{Hecht09}.  One of the motivations of the
current study is to examine the microscopic justification of the
tensor formulation of the Navier slip condition for surfaces with
anisotropic textures.

In this paper, the velocity fields and diffusion of fluid molecules
near interfaces between simple fluids and surfaces patterned with
anisotropic and random textures are studied by molecular dynamics
simulations.  For flows around parallel stripes of different
wettability, the transverse and longitudinal velocity components and
the effective slip length are compared against hydrodynamic
predictions. We will show that the directional diffusion coefficient
for fluid molecules in contact with wall atoms correlates well with
the effective slip length as a function of the flow direction with
respect to the stripe orientation.  In case of random wetting
patterns, the effective slip length depends sensitively on the total
area of wetting regions, in agreement with simple physical
arguments.

The rest of the paper is organized as follows. The details of
molecular dynamics simulations, parameter values, and thermostatting
procedure are described in the next section. The results for the
effective slip length at surfaces with periodic and random textures
and the numerical analysis of the interfacial diffusion of fluid
molecules are presented in Section~\ref{sec:Results}. The
conclusions are given in the last section.

\section{Details of molecular dynamics simulation model}
\label{sec:Model}

\begin{figure}[t]
\vspace*{-3mm}
\includegraphics[width=7.0cm,angle=0]{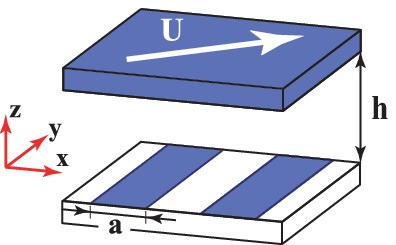}
\caption{(Color online) A schematic representation of the channel
geometry and surface patterns indicated by the blue color (wetting
regions) and white color (nonwetting regions). Steady shear flow is
induced by the upper wall moving with a constant velocity $U$ in the
$xy$ plane at an angle $\theta$ with respect to the $\hat{x}$
direction. The lower patterned wall is stationary.}
\label{fig:schematic}
\end{figure}

The schematic setup of the channel geometry and the orientation of
wetting and nonwetting regions of the stationary lower wall are
illustrated in Figure\,\ref{fig:schematic}. The steady shear flow is
induced by the upper wall moving with a constant velocity in the
$xy$ plane. The structure of three-dimensional flow fields in the
confined fluid is determined by the local boundary conditions at the
patterned lower wall and the orientation of the upper wall velocity
with respect to the stripe direction.

The pair interaction between fluid monomers ($N_{f}\!=6000$) is
modeled via the truncated Lennard-Jones (LJ) potential
\begin{equation}
V_{LJ}(r)\!=4\,\varepsilon\,\Big[\Big(\frac{\sigma}{r}\Big)^{12}\!-\Big(\frac{\sigma}{r}\Big)^{6}\,\Big],
\label{LJ}
\end{equation}
where $\varepsilon$ and $\sigma$ are the energy and length scales of
the fluid phase, and the cutoff radius is $r_c\!=\!2.5\,\sigma$.
Wall atoms interact with fluid monomers through a modified LJ
potential with adjustable strength of the attractive term
\begin{equation}
\widetilde{V}_{LJ}(r)\!=4\,\varepsilon\,\Big[\Big(\frac{\sigma}{r}\Big)^{12}\!-\delta\Big(\frac{\sigma}{r}\Big)^{6}\,\Big],
\label{LJw}
\end{equation}
where the parameter $\delta\,{=}\,1.0$ for wetting regions and
$\delta\,{=}\,0.1$ for nonwetting regions and the rest of the
parameters are the same, i.e., $\varepsilon_{\rm wf}=\varepsilon$,
$\sigma_{\rm wf}\,{=}\,\sigma$ and $r_c\!=\!2.5\,\sigma$. The wall
atoms do not interact with each other. For the results presented in
the next section, the parameter $\delta\,{=}\,1.0$ was fixed for the
upper wall atoms, while the lower wall is either homogeneous
($\delta\,{=}\,1.0$ or $\delta\,{=}\,0.1$) or patterned with
periodic stripes or random wetting regions.

\begin{figure}[t]
\vspace*{-3mm}
\includegraphics[width=9.0cm,angle=0]{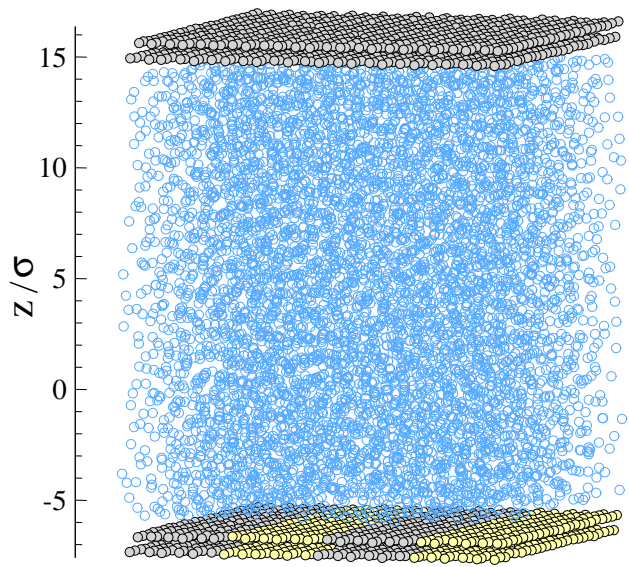}
\caption{(Color online) A snapshot of $6000$ fluid monomers (open
blue circles) confined between atomistic walls patterned with
wetting (filled gray circles) and nonwetting (filled yellow circles)
regions. The width of stripes at the lower stationary wall is
$a=4.42\,\sigma$ and the upper wall velocity is
$U=0.1\,\sigma/\tau$.}
\label{fig:snapshot}
\end{figure}

The solid walls are constructed of two layers of the face-centered
cubic (fcc) lattice with density $\rho_w\,{=}\,2.3\,\sigma^{-3}$.
Each layer is composed of $576$ lattice sites arranged on the
$(111)$ plane with $[11\bar{2}]$ orientation parallel to the
$\hat{x}$ direction. The nearest-neighbor distance between lattice
sites within the $(111)$ plane is $0.85\,\sigma$. The wall atoms are
attached to the lattice sites by harmonic springs. The system
dimensions in the $xy$ plane (parallel to the confining walls) were
kept fixed $L_x\,{=}\,17.67\,\sigma$ and $L_y\,{=}\,20.41\,\sigma$,
and the distance between wall lattice planes in contact with fluid
molecules was set $h\,{=}\,21.54\,\sigma$. The fluid density is
defined as a ratio of the total number of fluid monomers to the
volume accessible to the fluid phase
$\rho\,{=}\,N_{f}/L_{x}L_{y}(h-\sigma)\,{=}\,0.81\,\sigma^{-3}$.
Periodic boundary conditions for fluid monomers and wall atoms were
imposed along the $\hat{x}$ and $\hat{y}$ directions. The motion of
the upper wall with a constant velocity oriented at an angle
$\theta$ with respect to the $\hat{x}$ axis was modelled by
translating the fcc lattice sites and applying periodic boundary
conditions in the $xy$ plane at each time step.
Figure\,\ref{fig:snapshot} shows a snapshot of the fluid phase
confined between atomically smooth walls. In this particular case,
the upper wall velocity is oriented perpendicular to the stripe
direction.

The motion of wall atoms was coupled to an external heat bath by
adding Langevin noise and friction terms to all three components of
the equations of motion, e.g., in the $\hat{x}$ direction the
equation is given by
\begin{eqnarray}
\label{Langevin_wall_x} m_w\,\ddot{x}_i + m_w\,\Gamma\dot{x}_i & = &
-\sum_{i \neq j} \frac{\partial \widetilde{V}_{LJ\,ij}}{\partial
x_i} - \frac{\partial V_{sp}}{\partial x_i} + f_i\,,
\end{eqnarray}
where the mass of a wall atom is $m_w\,{=}\,100\,m$, the friction
coefficient is $\Gamma\,{=}\,2.0\,\tau^{-1}$, and $f_i$ is a random
force with zero mean and variance $\langle
f_i(0)f_j(t)\rangle\,{=}\,\,2mk_BT\Gamma\delta(t)\delta_{ij}$
determined from the fluctuation-dissipation theorem. The temperature
of the Langevin thermostat is $T\,{=}\,1.1\,\varepsilon/k_B$, where
$k_B$ is the Boltzmann constant. The wall atoms were tethered to the
fcc lattice sites under the harmonic potential
$V_{sp}\,{=}\,\frac{1}{2}\,\kappa\,r^2$ with the spring stiffness
coefficient $\kappa\,{=}\,2000\,\varepsilon/\sigma^2$. It was
previously shown that a sufficiently large value of the stiffness
coefficient does not affect the slip length at the interface between
monatomic fluids and dense crystalline walls~\cite{PriezjevJCP}. The
equations of motion for wall atoms and fluid monomers were solved
using the fifth-order Gear predictor-corrector scheme~\cite{Allen87}
with a time step $\triangle t\,{=}\,0.005\,\tau$, where
$\tau\!=\!\sqrt{m\sigma^2/\varepsilon}$ is the LJ time scale. Thus,
the oscillation time $2\pi\sqrt{m_w/\kappa}\approx1.4\,\tau$ of wall
atoms is much larger than the integration time step. Typical values
for liquid argon are $\sigma=0.34\,\text{nm}$,
$\varepsilon/k_{B}=120\,\text{K}$ and
$\tau=2.16\times10^{-12}\,\text{s}$~\cite{Allen87}.

A common practice in non-equilibrium MD simulations is to apply the
Langevin thermostat only in the direction of motion perpendicular to
the plane of shear in order to maintain constant temperature of the
fluid phase~\cite{Thompson90,Nature97}. It is expected, however,
that if the orientation of the upper wall velocity is restricted to
$0^{\circ}<\theta<90^{\circ}$ in the system shown in
Fig.\,\ref{fig:schematic}, then the fluid flow near the lower
patterned wall will have non-zero components of the averaged
velocity fields in all spatial dimensions, and, therefore, the
Langevin friction term applied to fluid monomers might bias the flow
profile. To test the dependence of our results on the thermostatting
procedure, we performed simulations with $a=8.84\,\sigma$ and the
upper wall velocity $U\,{=}\,\,0.1\,\sigma/\tau$ oriented at
$\theta\,{=}\,45^{\circ}$ relative to the $\hat{x}$ direction for
two cases where the Langevin thermostat with
$\Gamma\,{=}\,1.0\,\tau^{-1}$ was applied to the equations of motion
for fluid monomers in the direction either perpendicular to the
plane of shear or parallel to the $\hat{z}$ direction. After
averaging over thermal fluctuations, we observed a slight difference
in the velocity profiles near the lower wall which resulted in a
discrepancy between slip lengths of about $0.5\,\sigma$. To
eliminate the uncertainty associated with the friction term, in the
present study, the Langevin thermostat was applied only to the
equations of motion for wall atoms. The viscous heat in the fluid
phase was efficiently removed via interaction of fluid monomers with
thermal wall atoms so that the fluid temperature remained constant
$T_f\,{=}\,(1.10\pm0.01)\,\,\varepsilon/k_B$ even at the highest
upper wall velocity $U\,{=}\,\,0.1\,\sigma/\tau$. A similar
thermostatting procedure in sheared fluids confined by thermal walls
was implemented in previous MD studies~\cite{dePablo97,Freund11}.

In the present study, the simulations were performed at relatively
low shear rates $\dot{\gamma}\lesssim0.005\,\tau^{-1}$, which
required averaging of the velocity profiles over long time intervals
(up to $10^6\,\tau$) within horizontal bins of thickness $\Delta
z\!=0.01\,\sigma$. Likewise, the fluid density profiles were
computed within narrow bins of thickness $\Delta z\!=0.01\,\sigma$
to resolve fine details of the layered structure near interfaces.
The structure factor was computed in the first fluid layer according
to
$S(\mathbf{k})=|\sum_{j=1}^{N_{\ell}}e^{i\,\mathbf{k}\cdot\mathbf{r}_j}|^2/\,N_{\ell}$,
where $\mathbf{k}$ is a two-dimensional wave vector,
$\mathbf{r}_j\,{=}\,(x_j,y_j)$ is the position of the $j$th monomer,
and $N_{\ell}$ is the total number of monomers within the
layer~\cite{Thompson90}. The fluid viscosity
$\mu\,{=}\,(2.15\pm0.15)\,\,\varepsilon\tau\sigma^{-3}$ was
previously found to be rate independent for
$\dot{\gamma}\lesssim0.072\,\tau^{-1}$ and insensitive to the
temperature variation in the range of $1.1\leqslant
T\,k_B/\varepsilon\leqslant1.35$~\cite{Niavarani10}. An estimate of
the maximum value of the Reynolds number based on
$U\,{=}\,\,0.1\,\sigma/\tau$ and no-slip boundary conditions is
$Re=\rho\,\!h\,U/\mu\approx0.77$, which is clearly indicative of
laminar flow in the channel.

\section{Results}
\label{sec:Results}

\subsection{Fluid density and velocity profiles for homogeneous walls}
\label{subsection:homogeneous}

\begin{figure}[t]
\includegraphics[width=12.cm,angle=0]{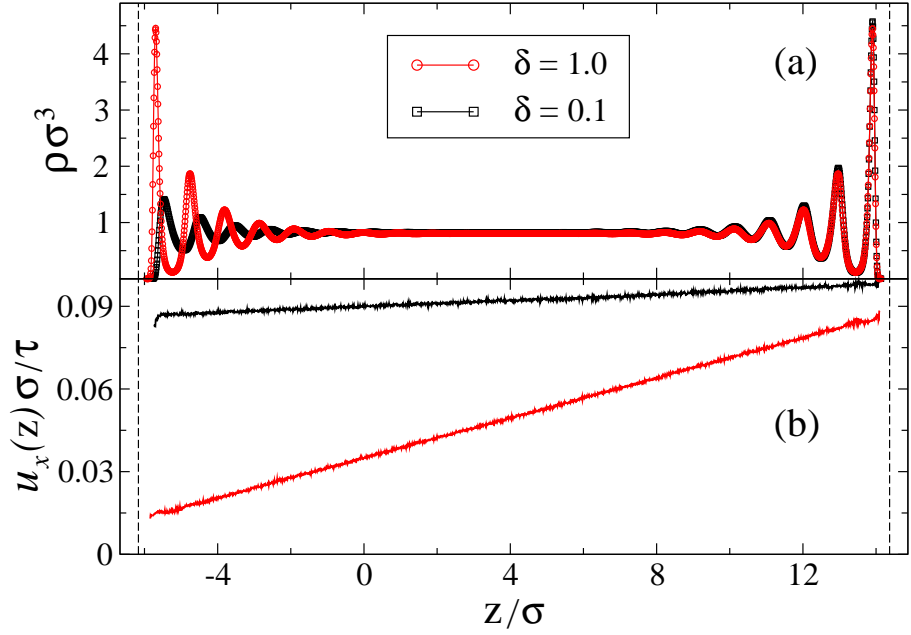}
\caption{(Color online) Averaged (a) density and (b) velocity
profiles across the channel for wetting $\delta\,{=}\,1$ (red
circles) and nonwetting $\delta\,{=}\,0.1$ (black squares) lower
walls. The upper wall velocity in the $\hat{x}$ direction is
$U\,{=}\,\,0.1\,\sigma/\tau$. The vertical axes at
$z/\sigma\,{=}\,-6.67$ and $14.87$ coincide with the location of the
fcc lattice planes in contact with fluid molecules. The vertical
dashed lines at $z/\sigma\,{=}\,-6.17$ and $14.37$ indicate the
location of liquid-solid interfaces.}
\label{fig:dens_velo_wet_nonwet}
\end{figure}

We first consider steady shear flow in the channel with a
homogeneous lower wall, which is either wetting ($\delta\,{=}\,1.0$)
or nonwetting ($\delta\,{=}\,0.1$). It is well known that the
presence of a flat crystalline surface promotes the formation of a
layered structure in the adjacent fluid, e.g., see
Ref.\,\cite{Kaplan06}. An example of fluid density profiles is shown
in Fig.\,\ref{fig:dens_velo_wet_nonwet}\,(a). As is evident, the
fluid layering is most pronounced near interfaces, the amplitude of
density oscillations gradually decays on distances of about five
molecular diameters away from the solid walls, and the fluid density
is uniform in the middle of the channel.

The interaction between fluid monomers and wall atoms is controlled
by the strength of the attractive term in the LJ potential. When the
parameter $\delta$ in Eq.\,(\ref{LJw}) is reduced, then the well
depth of the potential function decreases and the pairwise
separation, where the potential energy reaches a minimum value,
increases. For comparison, the minimum of the full LJ potential with
$\delta\,{=}\,1.0$ in Eq.\,(\ref{LJw}) occurs at
$r\!=\sqrt[6]{2}\,\sigma\approx1.12\,\sigma$ and equals
$\widetilde{V}_{LJ}(1.12\,\sigma)\!=-\varepsilon$, while the
modified LJ potential with $\delta\,{=}\,0.1$ has a much lower well
depth $\widetilde{V}_{LJ}(1.65\,\sigma)\!=-0.01\,\varepsilon$.
Therefore, it is not surprising that the amplitude of the first
fluid layer near the nonwetting lower wall in
Fig.\,\ref{fig:dens_velo_wet_nonwet}\,(a) is significantly reduced
and its location is shifted away from the wall. Note also that, due
to a pronounced fluid layering near the wetting lower wall, the
amplitude of density oscillations near the upper wall is slightly
smaller in the wetting case. Finally, we have checked that the
density profiles reported in
Fig.\,\ref{fig:dens_velo_wet_nonwet}\,(a) for the upper wall
velocity $U\,{=}\,\,0.1\,\sigma/\tau$ are the same as those computed
in the absence of shear flow (not shown).

The averaged fluid velocity profiles for wetting and nonwetting
lower walls are presented in
Fig.\,\ref{fig:dens_velo_wet_nonwet}\,(b). In both cases, the
velocity profiles are linear across the channel; however, the slip
velocity is much higher near the nonwetting lower wall. A slight
downward curvature in the velocity profile near the nonwetting
surface might be related to the fact that only fluid monomers with
relatively large velocity component in the $\hat{z}$ direction can
penetrate more deeply into the solid wall and, thus, the tangential
velocity in that region is computed from biased velocity
distribution. To compute the slip length, we define the location of
liquid-solid interfaces (see vertical dashed lines in
Fig.\,\ref{fig:dens_velo_wet_nonwet}) at the distance $0.5\,\sigma$
away from the fcc wall lattice planes in contact with fluid
monomers. For flows over homogeneous walls, the slip length is
determined from the linear fit to the velocity profiles excluding
regions of about $2\,\sigma$ near interfaces. Depending on the
wall-fluid interaction, the slip lengths are
$b_w=(3.6\pm0.8)\,\sigma$ for wetting surfaces and
$b_n=(156\pm10)\,\sigma$ for nonwetting surfaces. Within the
reported error bars, the slip lengths are independent of the shear
flow orientation relative to the fcc wall lattice. It was shown that
at low shear rates $\dot{\gamma}\lesssim0.005\,\tau^{-1}$ considered
in the present study, the slip length at an interface between smooth
crystalline walls and monatomic fluids is rate
independent~\cite{Nature97}.

\subsection{Anisotropic slip lengths for periodically patterned walls}
\label{subsection:stripes}

We next study the effects of shear flow orientation and stripe width
on the flow structure in the channel with the lower stationary wall
patterned with alternating stripes of different wettability, as
shown in Fig.\,\ref{fig:schematic}. In this geometry, the textured
lower wall induces wavy perturbations in simple shear flow which
penetrate into the fluid domain on a length scale of about a stripe
width, and the slip velocity at the lower wall is not parallel to
the upper wall velocity when $0^{\circ}<\theta<90^{\circ}$. In this
section, we only consider alternating stripes of equal width, which
are measured $a=L_{x}/\,n$, where integer $n =
2,~4,~8,~12,~\text{and}~24$. Thus, in all cases examined, the stripe
width ($a/\sigma\!=8.84,~4.42,~2.21,~1.47,~0.74$) is smaller than
the channel height $h\,{=}\,21.54\,\sigma$; and, therefore, the
longitudinal velocity profiles are expected to be linear across the
channel except in the region of about $a$ near the lower wall. In
what follows, we denote the longitudinal component of the fluid
velocity profile (parallel to the upper wall velocity) by
$u_{||}(z)$ and the transverse component of the velocity profile by
$u_{\perp}(z)$, which is perpendicular to the direction of $U$.

The problem of anisotropic slip flow over an array of periodic
stripes of mixed wettability was addressed analytically assuming
that the stripe width is much smaller that the fluid
domain~\cite{Wang03,Vinograd08,Teo09,Vinograd10}. In particular, it
was shown that the angular dependence of the effective slip length
is given by
\begin{equation}
L_s(\theta) = b_{\perp}cos^2\theta + b_{\parallel} sin^2\theta,
\label{BazantVinograd}
\end{equation}
where $b_{\perp}$ and $b_{||}$ are respectively slip lengths for
flows perpendicular ($\theta=0^{\circ}$) and parallel
($\theta=90^{\circ}$) to the stripe
orientation~\cite{Wang03,Vinograd10}. If the local slip lengths at
wetting and nonwetting regions are finite and independent of the
flow direction, then $b_{||}>b_{\perp}$ and in the special case of
stick-perfect slip stripes
$b_{||}=2\,b_{\perp}$~\cite{Lauga03,Vinograd10}. From the solution
of the Stokes equation with mixed boundary
conditions~\cite{Teo09,Vinograd10}, the ratio of slip velocities in
longitudinal and transverse directions can be calculated as a
function of the flow orientation
\begin{equation}
\frac{u_{\perp}^s}{u_{||}^s}=\frac{(b_{||}-b_{\perp})\,sin\,\theta\,cos\,\theta}
{b_{\perp} cos^2\theta + b_{\parallel} sin^2\theta}.
\label{uperp_upara}
\end{equation}
In the limiting cases when the flow direction is perpendicular or
parallel to the stripe orientation or when the surface is
homogeneous (i.e., $b_{||}=b_{\perp}$), the transverse slip velocity
in Eq.\,(\ref{uperp_upara}) vanishes. As an aside, the continuum
analysis also predicts that the transverse velocity component is
maximum when $\theta=45^{\circ}$~\cite{Teo09}.

\begin{figure}[t]
\includegraphics[width=12.cm,angle=0]{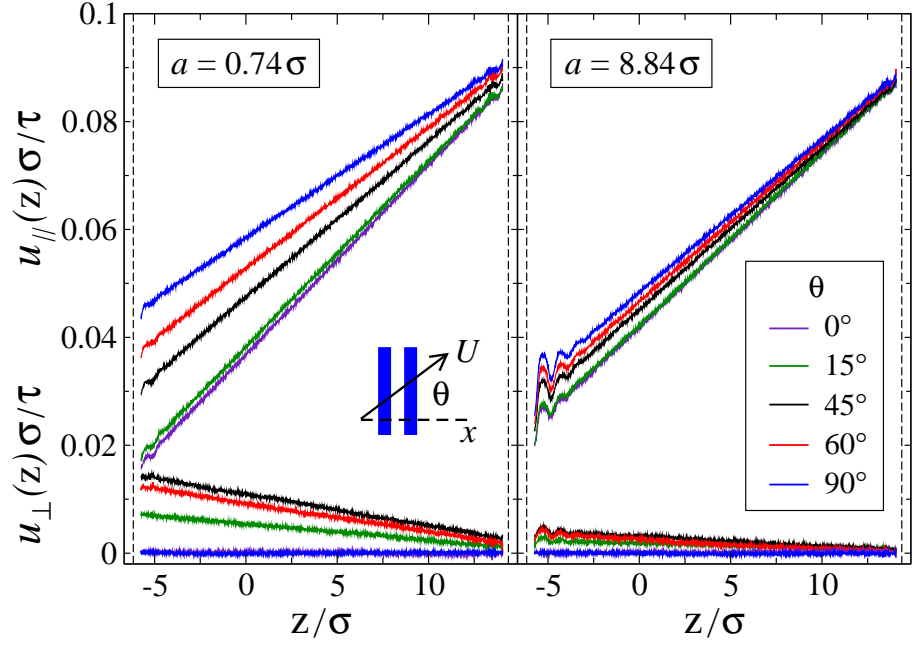}
\caption{(Color online) Averaged longitudinal (upper set of curves)
and transverse (lower set of curves) velocity profiles for the
indicated values of $\theta$ and stripe widths $a=0.74\,\sigma$
(left panel) and $a=8.84\,\sigma$ (right panel). The vertical axes
coincide with the location of the fcc lattice planes at
$z/\sigma\,{=}\,-6.67$ and $14.87$. The vertical dashed lines at
$z/\sigma\,{=}\,-6.17$ and $14.37$ indicate liquid-solid
interfaces.} \label{fig:velo_pattern}
\end{figure}

Examples of longitudinal and transverse velocity profiles for the
smallest ($a=0.74\,\sigma$) and largest ($a=8.84\,\sigma$) stripe
widths are presented in Fig.\,\ref{fig:velo_pattern} for selected
values of $\theta$. In both cases, the longitudinal velocity
component is maximum (minimum) when the upper wall velocity is
parallel (perpendicular) to the stripe orientation, and the
transverse flow is maximum when $\theta=45^{\circ}$, in agreement
with the continuum analysis~\cite{Teo09,Vinograd10}. If the stripe
width is about the molecular size, the alternating surface potential
[\,$\delta\,{=}\,1.0~\text{and}~0.1$ in Eq.\,(\ref{LJw})\,]
represents an effectively roughened surface for the flow component
perpendicular to the stripe orientation. As a consequence, the
location of the first fluid layer varies periodically above wetting
and nonwetting regions [\,e.g., see
Fig.\,\ref{fig:dens_velo_wet_nonwet}\,(a)\,], and the slip velocity
along the $\hat{x}$ direction (perpendicular to stripes) is reduced.
In contrast, when the flow is parallel to stripes, fluid monomers
are transported along homogeneous wetting or nonwetting regions, and
the effect of surface roughness is absent. This explains the
relatively large variation of the longitudinal slip velocity as a
function of the flow orientation for the stripe width
$a=0.74\,\sigma$. For the largest stripe width $a=8.84\,\sigma$, the
velocity profiles acquire pronounced oscillations near the lower
wall because of the mismatch between the location of peaks in
density profiles above wetting and nonwetting regions [\,shown in
Fig.\,\ref{fig:dens_velo_wet_nonwet}\,(a)\,]. Similar effects were
reported in the previous study where only flows parallel and
perpendicular to stripes were considered~\cite{Priezjev05}. For the
results presented below, the effective slip length was computed by
extrapolating the linear part of the longitudinal velocity profiles
to zero velocity.

\begin{figure}[t]
\includegraphics[width=12.cm,angle=0]{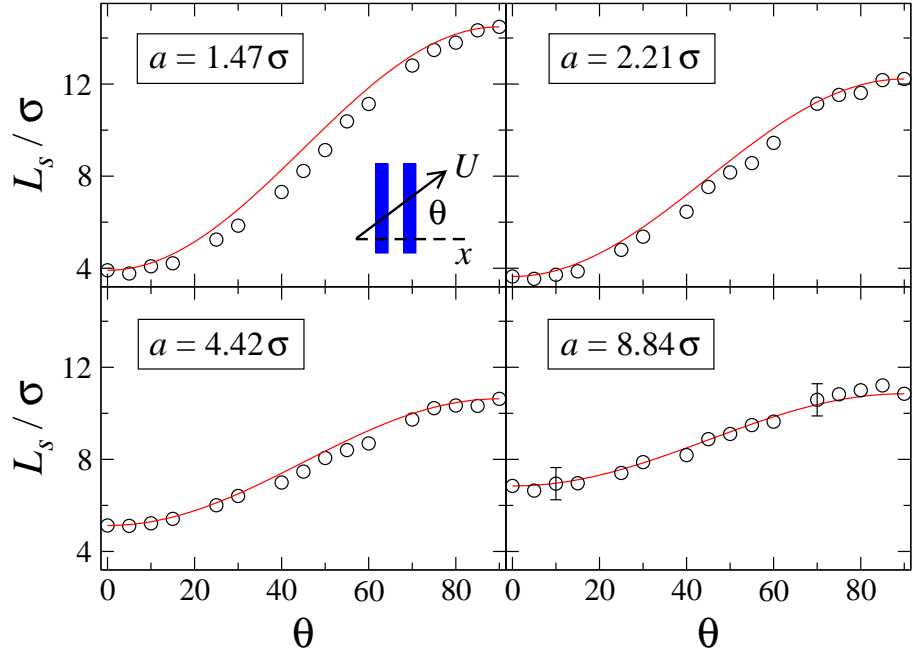}
\caption{(Color online) Variation of the effective slip length
$L_s/\sigma$ as a function of $\theta$ for the indicated values of
the stipe width. The MD data are shown by open circles. The red
curves are hydrodynamic predictions computed using
Eq.\,(\ref{BazantVinograd}).} \label{fig:ls_theta}
\end{figure}

The angular dependence of the effective slip length for the
indicated values of the stripe width is presented in
Fig.\,\ref{fig:ls_theta}. As expected, $L_s$ monotonically increases
as $\theta$ approaches $90^{\circ}$. For a given stripe width, the
values $b_{\perp}$ and $b_{||}$ were determined from the
longitudinal velocity profiles and then used in
Eq.\,(\ref{BazantVinograd}) to compare the results of MD simulations
with continuum predictions (shown by red curves in
Fig.\,\ref{fig:ls_theta}). The agreement between the MD data and
continuum solution Eq.\,(\ref{BazantVinograd}) becomes progressively
better as the stripe width increases up to $a=8.84\,\sigma$. It
should be mentioned that these results do not contradict the
conclusion drawn from the previous study~\cite{Priezjev05}, which
demonstrated that the agreement between continuum analysis and MD
simulations holds when the stripe width is larger than about thirty
molecular diameters. In the context of the present study, this
conclusion could be verified by computing $L_s(\theta)$ directly
from the solution of the Stokes equation for the flow geometry shown
in Fig.\,\ref{fig:schematic} with the local slip lengths at wetting
and nonwetting regions extracted from MD simulations. Such an
analysis was not performed in the current study.

\begin{figure}[t]
\includegraphics[width=12.cm,angle=0]{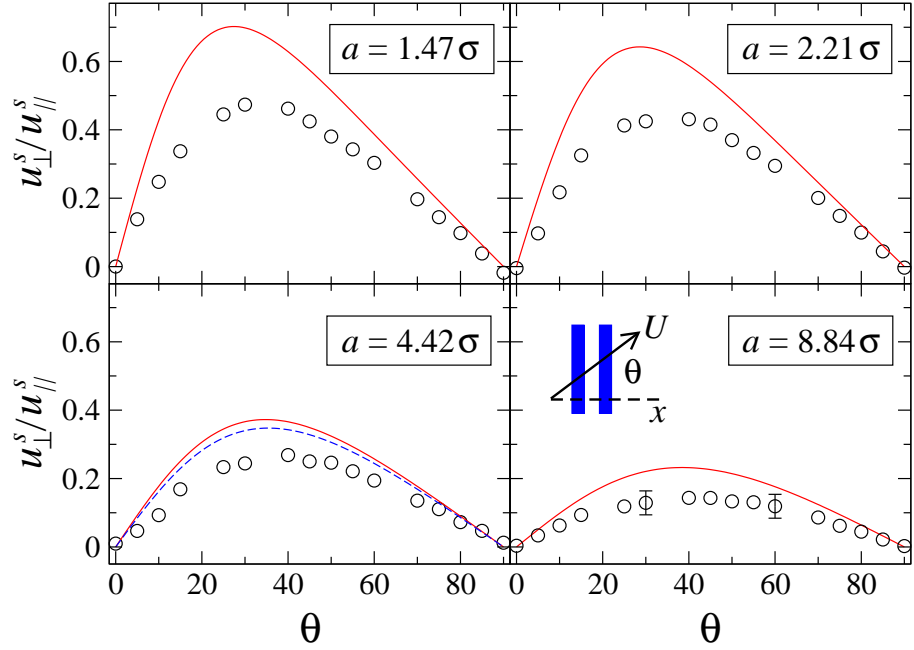}
\caption{(Color online) The ratio of transverse and longitudinal
components of the slip velocity as a function of $\theta$. The MD
data are shown by open circles. The solid red curves are continuum
predictions calculated using Eq.\,(\ref{uperp_upara}). The dashed
blue curve is Eq.\,(\ref{uperp_upara}) where $b_{\perp,||}$ are
defined with respect to the location of the first fluid layer (see
text for details).}   \label{fig:us_theta}
\end{figure}

\begin{figure}[t]
\includegraphics[width=12.cm,angle=0]{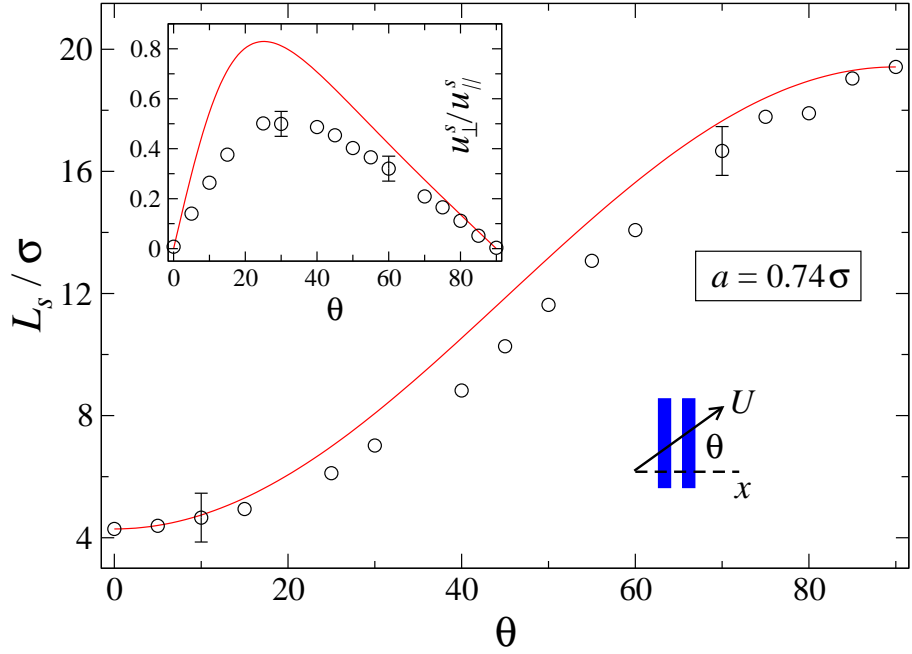}
\caption{(Color online) The effective slip length $L_s/\sigma$ as a
function of $\theta$ for the stripe width $a=0.74\,\sigma$. The
continuum prediction, Eq.\,(\ref{BazantVinograd}), is shown by the
red curve. The inset shows the angular dependence of the ratio of
slip velocities in transverse and longitudinal directions. The red
curve in the inset is the prediction of Eq.\,(\ref{uperp_upara}).}
\label{fig:ls_theta600}
\end{figure}

As discussed above, the transverse flow appears when the upper wall
velocity is neither parallel nor perpendicular to the stripe
orientation. Next, we present a more detailed comparative analysis
of the transverse slip velocity at the lower wall based on the MD
data and the continuum solution Eq.\,(\ref{uperp_upara}). In MD
simulations, the longitudinal and transverse slip velocity
components of the first fluid layer were computed as follows:
\begin{equation}
u_{\perp,||}^{s}=\int_{z_0}^{z_1}\!u_{\perp,||}(z)\,\rho(z)\,dz
\,\Big/ \int_{z_0}^{z_1}\!\rho(z)\,dz, \label{velo_defin}
\end{equation}
where the limits of integration ($z_0=-5.1\,\sigma$ and
$z_1=-5.8\,\sigma$) are defined by the width of the first peak in
the density profile above the patterned lower wall. The ratio
$u_{\perp}^{s}/u_{||}^{s}$ as a function of $\theta$ is plotted in
Fig.\,\ref{fig:us_theta}. The results show that although the
magnitude of the transverse slip velocity is largest when
$\theta=45^{\circ}$, the maximum angle between the upper wall
velocity and the slip velocity occurs when $\theta<45^{\circ}$. For
example, this angle is about $25^{\circ}$ for the stripe width
$a=1.47\,\sigma$ when $\theta=30^{\circ}$. Further, the MD values
$b_{\perp}(a)$ and $b_{||}(a)$ were used in Eq.\,(\ref{uperp_upara})
to compute the ratio $u_{\perp}^{s}/u_{||}^{s}$ as a function of
$\theta$ (see red curves in Fig.\,\ref{fig:us_theta}). The continuum
solution Eq.\,(\ref{uperp_upara}) is only in qualitative agreement
with the MD data. The discrepancy might be attributed to the
roughness effect discussed earlier and to the fact that the stripe
width is comparable to the fluid molecular size. Another possible
contributing factor is the uncertainty in defining the exact
location of the liquid-solid interface used for computing the
effective slip length (e.g., see the vertical dashed lines in
Fig.\,\ref{fig:velo_pattern}). Remember that the position of the
first fluid layer is displaced by about $\sigma$ from the fcc
lattice plane. Thus, if the location of the interface is taken at
the position of the first fluid layer, then the effective slip
lengths become $b_{\perp,||}\rightarrow b_{\perp,||}+0.5\,\sigma$.
The corresponding continuum solution Eq.\,(\ref{uperp_upara}) is
shown by the blue dashed curve in Fig.\,\ref{fig:us_theta} when the
stripe width is $a=4.42\,\sigma$. Notice the small difference
between continuum solutions, which, in general, is expected to be
negligible when $b_{\perp,||}\gg0.5\,\sigma$. For completeness, the
results for the smallest stripe width $a=0.74\,\sigma$ are reported
in Fig.\,\ref{fig:ls_theta600}. In this case, the discrepancy
between the MD data and continuum solutions
Eqs.\,(\ref{BazantVinograd})-(\ref{uperp_upara}) is most pronounced.

\subsection{Interfacial diffusion near surfaces of patterned wettability}
\label{subsection:diffusion_stripes}

A tensorial generalization of the Navier slip condition for flows
over anisotropic surfaces involves a relation between the normal
traction at the interface and fluid slip velocity via an interfacial
mobility tensor~\cite{Vinograd08}. In analogy with the theory of
Brownian motion, it was conjectured that the mobility of fluid
molecules near anisotropic surfaces is directly related to the
interfacial diffusivity per unit area~\cite{Vinograd08}. Simply put,
it implies that the effective slip length at the interface between a
Newtonian fluid and a textured surface is proportional to the
diffusion coefficient of fluid molecules near the surface. In the
present study, the diffusion coefficient was estimated from
two-dimensional trajectories of fluid monomers within the first
layer near the patterned wall at equilibrium (i.e., when both walls
are at rest).

The numerical analysis of the molecular displacement was performed
only for those fluid monomers that remained in contact with the
lower wall atoms (in the first fluid layer) during the diffusion
time interval. It should be noted that it is important not to
include fluid monomers further away from the patterned wall because
their diffusion in the $xy$ plane quickly becomes isotropic. For
example, it was recently shown that when the LJ fluid is confined in
narrow slit-pores (with the channel height of about $5\,\sigma$) and
both walls are patterned with stripes of different wettability, then
the mean square displacement curves, which were averaged over all
fluid molecules in the direction either parallel or perpendicular to
stripes, nearly coincide with each other~\cite{Gubbins07}.

\begin{figure}[t]
\includegraphics[width=12.cm,angle=0]{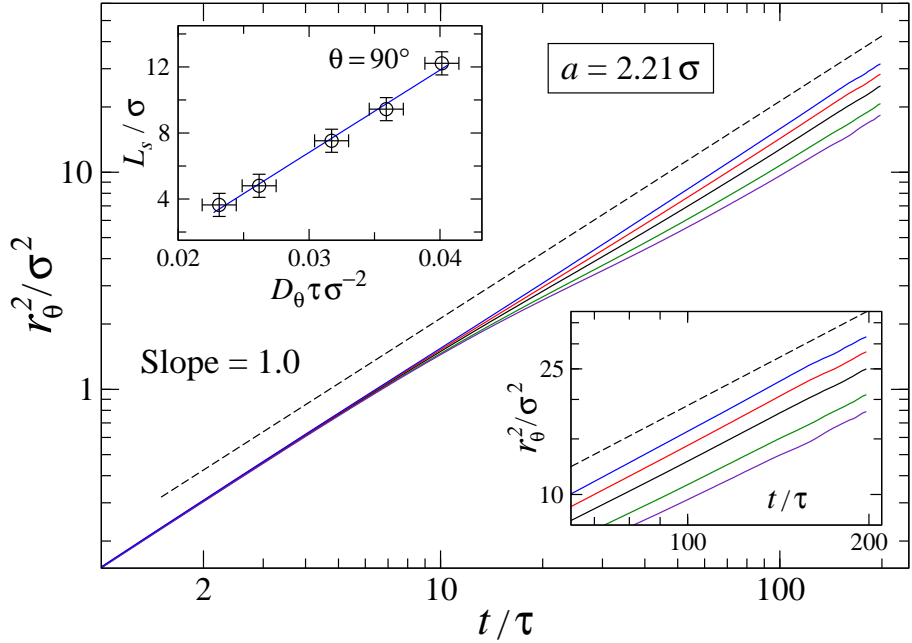}
\caption{(Color online) The mean square displacement of fluid
monomers in the first layer near the lower patterned wall with the
stripe width $a=2.21\,\sigma$, when both walls are at rest. Each
curve is computed by taking the component of the displacement vector
along a line oriented at an angle
$\theta=0^{\circ},~25^{\circ},~45^{\circ},~60^{\circ},~90^{\circ}$
with respect to the $\hat{x}$ axis (from bottom to top). The dashed
line with unit slope is plotted for reference. The lower inset shows
an expanded view of the same data at large $t$. The upper inset
shows a correlation between the in-plane diffusion coefficient and
the effective slip length for the same set of $\theta$. The straight
blue line is the best fit to the data.} \label{fig:msd_diff}
\end{figure}

Figure\,\ref{fig:msd_diff} shows the time dependence of the mean
square displacement (MSD) of fluid monomers near the lower wall with
stripes of width $a=2.21\,\sigma$. For each curve plotted in
Fig.\,\ref{fig:msd_diff}, the displacement vector in the $xy$ plane
was projected onto a line oriented at an angle
$\theta=0^{\circ},~25^{\circ},~45^{\circ},~60^{\circ},~90^{\circ}$
with respect to the $\hat{x}$ axis. We find that the diffusion is
isotropic only when the average displacement of fluid monomers is
less than the stripe width. When $r_{\theta}\gtrsim a$, the mean
square displacement curves exhibit a gradual crossover to a linear
regime ($r^{2}_{\theta}\sim t$) where the diffusion becomes
anisotropic. Averaging over long time periods (up to
$2.6\times10^7\,\tau\approx5.6\times10^{-5}\,\text{s}$) was required
to accumulate good statistics because of the relatively low
probability that a fluid monomer would remain in the first layer for
a long time; especially above the nonwetting regions where the fluid
density layering is reduced and fluid monomers jump in and out of
the first layer more frequently.

The diffusion coefficient was computed from the linear slope of the
mean square displacement as a function of time
($r^2_{\theta}=4D_{\theta}\,t$) in the regime $t\gtrsim145\,\tau$,
which is shown in more detail in the lower inset of
Fig.\,\ref{fig:msd_diff}. The slight nonlinearity in MSD curves at
large times is reflected in the error bars for the diffusion
coefficient. Most interestingly, as shown in the upper inset of
Fig.\,\ref{fig:msd_diff}, there is an almost linear correlation
between the in-plane diffusion coefficient and the effective slip
length as a function of $\theta$. It means that the longitudinal
component of the slip velocity along the shear flow direction is
proportional to the diffusion rate of fluid monomers along that
direction at equilibrium. The results in Fig.\,\ref{fig:msd_diff}
provide support for the microscopic justification of the tensorial
formulation of the Navier slip boundary condition in the case of
Newtonian fluids and molecular-scale surface textures. A similar
correlation between $D_{\theta}$ and $L_s$ also holds for smaller
stripe widths ($a=0.74\,\sigma$ and $a=1.47\,\sigma$), although the
difference in diffusion rates along
$\theta=0^{\circ}~\text{and}~90^{\circ}$ directions becomes smaller
when the stripe width decreases (not shown). For larger stripe
widths ($a=4.42\,\sigma$ and $a=8.84\,\sigma$), an accurate
resolution of the mean square displacement curves in the linear
regime ($r_{\theta}>a$) would require a long averaging time because
of the low probability that a fluid monomer will remain within the
first layer for a long time interval.

\subsection{Slip flows over surfaces with random textures}
\label{subsection:random}

In general, the problem of slip flow over surfaces with mixed
boundary conditions specified on randomly distributed regions is
difficult to treat either analytically or numerically. An example of
a statistical analysis of the effective slip boundary condition for
liquid flow over a plane boundary with randomly distributed
free-slip regions was presented in Ref.\,\cite{Prosper07}. It was
found that the effective slip length is proportional to the typical
size of free-slip regions and a factor that depends on the
fractional area coverage~\cite{Prosper07}. In this section, we
consider a simple shear flow over a smooth substrate with random
distribution of wetting and nonwetting regions. The system setup is
essentially the same as described in Sec.\,\ref{sec:Model}, except
that the parameter $\delta$ in Eq.\,(\ref{LJw}) is randomly chosen
to be either $0.1$ or $1.0$ for the lower wall atoms. Each of two
fcc lattice layers of the lower wall contains the same number of
weakly ($\delta=0.1$) or strongly ($\delta=1.0$) attractive atoms.
In what follows, a fraction of wall atoms with $\delta\,{=}\,1.0$ is
denoted by $\phi$. Due to limited computational resources only one
realization of disorder was considered for each value of $\phi$.

\begin{figure}[t]
\includegraphics[width=12.cm,angle=0]{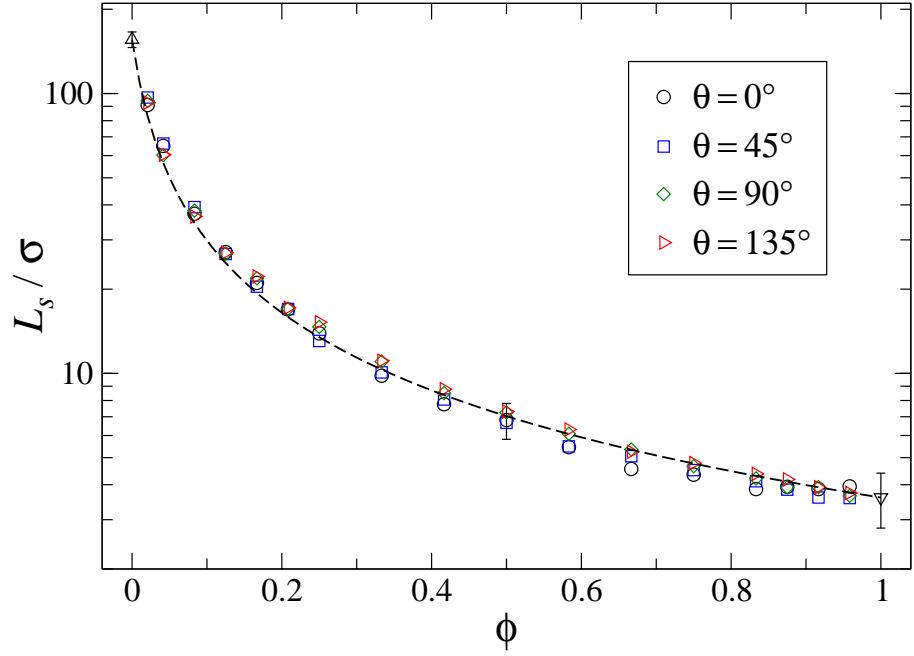}
\caption{(Color online) The effective slip length $L_s/\sigma$
versus the fraction of randomly distributed attractive atoms
($\delta=1.0$) at the lower wall. The orientation of the upper wall
velocity is specified in the inset. The black dashed curve is
Eq.\,(\ref{ls_random_eq}) with $b_w(\phi=1)=3.6\,\sigma$ and
$b_n(\phi=0)=156\,\sigma$.}   \label{fig:ls_phi}
\end{figure}

Figure\,\ref{fig:ls_phi} shows the effective slip length as a
function of $\phi$ for four orientations of the upper wall velocity
relative to the lower wall. As expected, $L_s$ decreases with
increasing the total area of wetting regions. Note that for each
value of $\phi$, the flow is almost isotropic; the slight
discrepancy in most probably due to finite size effects. When
$\phi=0.5$, the averaged effective slip length
$L_s\approx7.0\,\sigma$ is between $b_{\perp}$ and $b_{||}$ for any
value of the stripe width reported in Figs.\,\ref{fig:ls_theta} and
\ref{fig:ls_theta600}, which confirms earlier conclusions that
parallel (perpendicular) stripes attain maximum (minimum) slippage.
In the limiting cases $\phi\rightarrow0~\text{or}~\phi\rightarrow1$,
the fluid velocity fields near the lower wall are parallel to the
$xy$ plane, and the total friction coefficient (the ratio of fluid
viscosity to slip length) can be estimated by simply adding
contributions from wetting and nonwetting areas as follows:
\begin{equation}
\frac{\mu}{L_s(\phi)}=\frac{\mu\,\phi}{b_w}+\frac{\mu\,(1-\phi)}{b_n},
\label{add_friction}
\end{equation}
where $b_w$ and $b_n$ are respectively slip lengths for wetting
($\phi=1$) and nonwetting ($\phi=0$) surfaces. This immediately
gives the effective slip length as a function of $\phi$
\begin{equation}
L_s(\phi)=\frac{b_w\,b_n}{\phi\,b_n + (1-\phi)\,b_w}.
\label{ls_random_eq}
\end{equation}
As shown in Fig.\,\ref{fig:ls_phi}, the agreement between the MD
data and Eq.\,(\ref{ls_random_eq}) is quite good for all $\phi$ in
the range $[0,1]$. However, this correspondence might, in general,
not hold at intermediate values of $\phi$ and larger system size in
the $\hat{x}$ and $\hat{y}$ directions because of the spatial
variation of velocity profiles induced by the heterogeneous surface.
Interestingly, the formula for the effective friction coefficient,
Eq.\,(\ref{add_friction}), also accurately describes hydrodynamic
flows along alternating stripes with local slip lengths that are
larger than the system size~\cite{Bocquet04,Hendy07}. In addition,
it was shown numerically that an interpolation formula like
Eq.\,(\ref{add_friction}) predicts the effective slip length for
composite interfaces which consist of periodically distributed solid
and gas areas~\cite{BocqScale07}.

\begin{figure}[t]
\includegraphics[width=12.cm,angle=0]{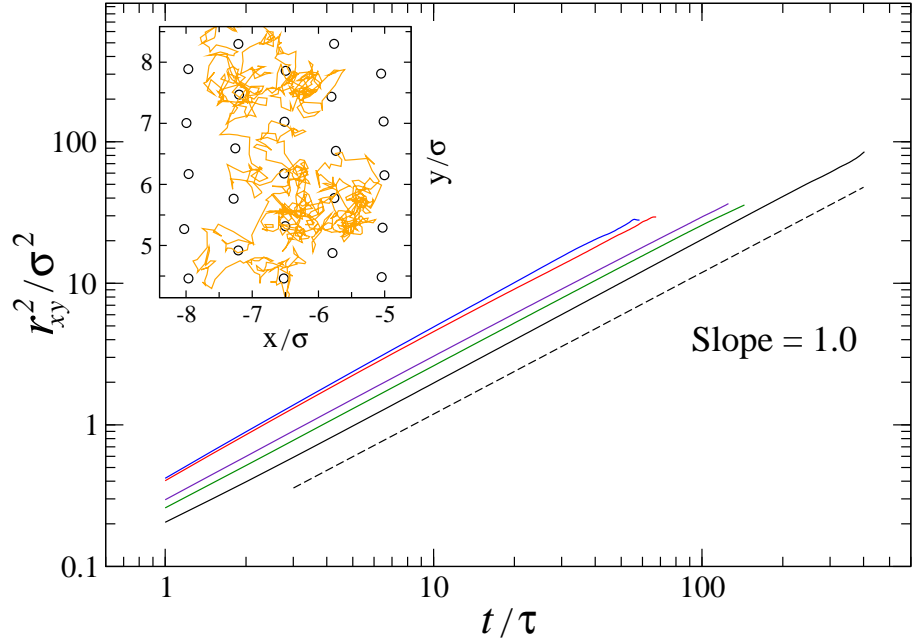}
\caption{(Color online) The mean square displacement of fluid
monomers in the first layer near the lower wall with
$\phi=0,~0.08,~0.58,~0.75,~1.0$ (from left to right) when $U=0$. The
dashed line with unit slope is plotted as a reference. The inset
shows a typical trajectory of a fluid monomer for about $100\,\tau$
near the wetting wall $\phi=1.0$. The positions of the fcc lattice
atoms in the $xy$ plane are denoted by open circles.}
\label{fig:msd_diff_rand}
\end{figure}

Similar to the analysis of the interfacial diffusion presented in
Sec.\,\ref{subsection:diffusion_stripes}, we evaluate the in-plane
diffusion coefficient in the absence of shear flow for fluid
monomers in the first layer near the lower wall for the same
realization of disorder as in Fig.\,\ref{fig:ls_phi}. As an example,
a typical trajectory projected onto the $xy$ plane is shown in the
inset of Fig.\,\ref{fig:msd_diff_rand}. It can be seen that when
$\phi=1$, the diffusive motion of a fluid monomer in contact with
lower wall atoms is strongly influenced by the periodic surface
potential; most of the time the monomer resides near the local
minima of the surface potential. In this case, the surface-induced
structure in the first fluid layer is quantified by a distinct peak
in the structure factor at the main reciprocal lattice vector
$S(8.53\,\sigma^{-1},0)\approx1.7$, which is comparable to the
height of a circular ridge $S(2\pi/\sigma)\approx2.9$ characteristic
of the short range order. As the fraction of wetting regions
decreases, the amplitude of density oscillations near the lower wall
is reduced [see Fig.\,\ref{fig:dens_velo_wet_nonwet}\,(a)],
trajectories of fluid monomers become less affected by the
corrugation of the random surface potential, and the time interval
between jumps in and out of the first fluid layer decreases. When
$\phi\lesssim0.08$, the in-plane structure factor, estimated in the
first fluid layer at equilibrium, does not contain any peaks at the
reciprocal lattice vectors.

\begin{figure}[t]
\includegraphics[width=12.cm,angle=0]{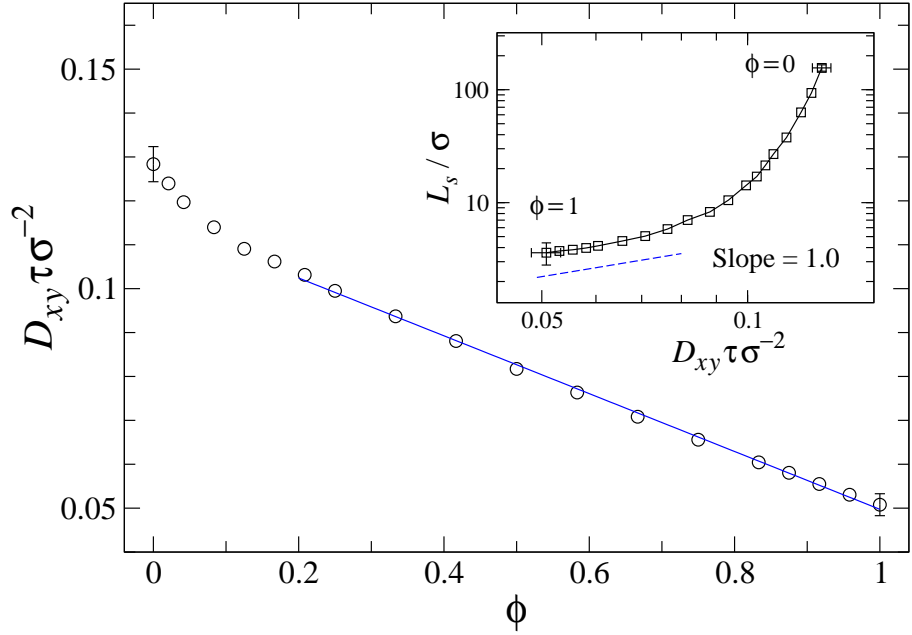}
\caption{(Color online) The in-plane diffusion coefficient $D_{xy}$
(in units $\sigma^2/\tau$) as a function of $\phi$. The straight
blue line is the best fit to the data for $\phi>0.2$. The inset
shows a correlation between the effective slip length and the
diffusion coefficient for $0\leqslant\phi\leqslant1$. The solid
curve is a guide for the eye. The dashed line with unit slope is
drawn for reference.}   \label{fig:diff_phi_slip_length}
\end{figure}

The mean square displacement curves as a function of time are
displayed in Fig.\,\ref{fig:msd_diff_rand} for selected values of
$\phi$. It is apparent that the diffusion becomes faster as the
total area of nonwetting regions increases. The in-plane diffusion
coefficient was estimated from the Einstein relation
$r^2_{xy}=4D_{xy}t$ when $t\gtrsim6\,\tau$. As shown in
Fig.\,\ref{fig:diff_phi_slip_length}, the diffusion coefficient
gradually varies between two values obtained for homogeneous
surfaces with $\phi=0$ and $\phi=1$. Furthermore, a correlation
between the effective slip length (averaged over four orientations
of the mean flow) and the in-plane diffusion coefficient is
presented in the inset Fig.\,\ref{fig:diff_phi_slip_length}. These
data indicate a nearly linear dependence between $L_s$ and $D_{xy}$
when the fraction of wetting regions is large; and, as discussed
earlier, there is a strong coupling between the diffusion of fluid
monomers in the first layer and the periodic surface potential. In
the opposite limit of small $\phi$, the net adsorption energy is
reduced and the in-plane diffusion of fluid monomers is mostly
dominated by the interaction with its fluid neighbors. In this
regime, the effective slip length increases rapidly as the fraction
of wetting regions decreases.

\section{Conclusions}

In this study, molecular dynamics simulations were performed in
order to investigate the effective slippage and molecular diffusion
at surfaces patterned with periodic or random textures.    In our
setup, the typical size of surface patterns is smaller than the
channel dimensions, and the local boundary conditions at homogeneous
wetting or nonwetting surfaces are described by finite slip lengths.
Particular attention was paid to the implementation of a
thermostatting procedure that does not bias flow profiles and
diffusion of fluid monomers.

For flows over surfaces patterned with stripes of different
wettability, the heterogeneous surfaces induce wavy perturbations in
velocity profiles and the slip velocity acquires a transverse
component. In this case, the effective slip length depends of the
shear flow direction with respect to the stripe orientation. We
found that the angular dependence of the effective slip length
computed by molecular dynamics simulations agrees well with the
analytical solution of the Stokes equation provided that the stripe
width is larger than several molecular diameters. At the same time,
however, the ratio of the transverse and longitudinal components of
the slip velocity agrees only qualitatively with hydrodynamic
predictions. Furthermore, the interfacial diffusion coefficient of
fluid molecules correlates well with the effective slip length as a
function of the shear flow direction.  The numerical analysis was
performed only for fluid monomers that remain in contact with the
wall atoms during the diffusion time interval. These findings lend
support for the microscopic justification of the tensor formulation
of the effective slip boundary conditions for noninertial flows of
Newtonian fluids over smooth surfaces with nanoscale anisotropic
textures.

In case of random surface textures, the simulation results and
simple physical arguments show that the effective slip length is
determined by the total area of wetting regions. When the fraction
of wetting regions is large, the diffusive motion of fluid monomers
is strongly influenced by the periodic surface potential; and the
effective slip length is nearly proportional to the in-plane
diffusion coefficient at equilibrium.  In the opposite limit of
small wetting areas, the diffusion of fluid monomers is less
affected by the corrugation of the surface potential, and the
effective slip length depends sensitively on the number of strongly
attractive wall atoms.

\section*{Acknowledgments}

Financial support from the National Science Foundation
(CBET-1033662) is gratefully acknowledged. The author would like to
thank G. Drazer for useful discussions. Computational work in
support of this research was performed at Michigan State
University's High Performance Computing Facility and the NNIN/C Nano
Computational Cluster at the University of Michigan.

\bibliographystyle{prsty}

\end{document}